\documentclass[11pt]{article}

\usepackage{geometry}
\usepackage{graphicx}
\usepackage{epstopdf}
\usepackage{footmisc}

\usepackage{natbib}

\usepackage{color}

\usepackage{amsmath}
\usepackage{amssymb}
\usepackage{empheq}
\usepackage{yhmath}

\usepackage{xspace}
\usepackage{subfig}
\usepackage{accents}

\def \e { \mbox{$\mathrm{e}$} }
\def \vth {v_{\mathrm{th}}}

\def \rhovec { \mbox{\boldmath $\rho$} }

\newcommand{\poiss}[2]{\{#1,\;#2\}}

\newcommand{\nAvg}[1]{\tilde{#1}}

\providecommand\bnabla{\boldsymbol{\nabla}}

\newcommand\eg{\textit{e.g.}\xspace}
\newcommand\ie{\textit{i.e.}\xspace}

\newcommand{\tAvg}[1]{\wideparen{#1}}
\newcommand{\sAvg}[1]{\left\langle#1\right\rangle_{s}}
\newcommand{\gyroavg}[1]{\left\langle#1\right\rangle_{\bf R}}
\newcommand{\angleavg}[1]{\left\langle#1\right\rangle_{\bf r}}
\newcommand{\CollisionOp}[1]{\gyroavg{C[#1]}}

\newlength{\dhatheight}

\usepackage{authblk}

\title{Enstrophy non-conservation and the forward cascade of energy in two-dimensional electrostatic magnetized plasma turbulence}
\author[1]{G.G. Plunk\thanks{gplunk@ipp.mpg.de}}
\affil[1]{Max Planck Institute for Plasma Physics, Greifswald 17491, Germany}

\begin{document}

\maketitle

\begin{abstract}
A fluid system is derived to describe electrostatic magnetized plasma turbulence at scales somewhat larger than the Larmor radius of a given species.  It is related to the Hasegawa-Mima equation, but does not conserve enstrophy, and, as a result, exhibits a forward cascade of energy, to small scales.  The inertial-range energy spectrum is argued to be shallower than a $-11/3$ power law, as compared to the $-5$ law of the Hasegawa-Mima enstrophy cascade.  This property, confirmed here by direct numerical simulations of the fluid system, may help explain the fluctuation spectrum observed in gyrokinetic simulations of streamer-dominated electron-temperature-gradient driven turbulence \citep{plunk-prl-2019}, and also possibly some cases of ion-temperature-gradient driven turbulence where zonal flows are suppressed \citep{plunk-prl-2017}.
\end{abstract}

\section{Introduction}

The turbulent cascade, a mechanism for the nonlinear transfer of energy across scales, is a key idea for understanding kinetic magnetized plasma turbulence.  By considering simplified models, in uniform magnetic geometry, one can obtain a theoretical prediction for the spectrum of fluctuations, valid across an ``inertial range'' of scales, free from energy sources and sinks.  Though such a theory is not able to fully describe the behavior of realistic turbulence, which hosts instabilities, damped modes, complicated magnetic geometries, {\em etc.}, it nevertheless constitutes a quantitative prediction of nonlinear behavior of the underlying gyrokinetic equation, an equation which generally governs actual systems of practical interest.  The existence of such theoretical test cases is valuable for validating the solution methods employed by gyrokinetic codes, and as a foundation for physical interpretation of the volumes of data they produce.

Here, a novel quasi-two-dimensional fluid system is derived from the electrostatic gyrokinetic system, to describe fluctuations that predominantly vary in the directions perpendicular to the mean magnetic field, \ie in the ``drift plane'', at scales $\ell$ larger than the Larmor radius $\rho$, corresponding to a species of interest.  The notion that quasi-two-dimensional behavior might underly magnetized plasma turbulence is intuitively justified by the fact that the magnetic guide field renders the dynamics inherently anisotropic.  Furthermore, instabilities that drive electrostatic turbulence in fusion plasmas, \eg the ion-temperature-gradient (ITG) and electron-temperature-gradient (ETG) modes, exhibit a kind of localization along the field line, accompanied by the domination of perpendicular dynamics over parallel dynamics.  The fluid limit $\ell \gg \rho$ is of particular importance, because the energy of the fluctuations is predominantly found at such scales -- these are the scales of importance, most directly affecting the performance of fusion devices.  Furthermore, the reduction of complexity afforded by fluid limits can reveal important features of the dynamics, that do not manifest in the analysis of the general gyrokinetic equations.

Although similar systems as the one presented here have been proposed and studied in the past, most notably the Hasegawa-Mima (HM) equation, \cite{hasegawa}, the present derivation takes special care in considering the consequences of the appearance of nonlinear finite-Larmor-radius (FLR) terms that appear in the dynamical equation for the electrostatic potential -- \ie the ``vorticity'' equation.  Such terms introduce a closure problem in the fluid moment hierarchy, where lower moments are coupled to ever higher ones, generally without end.  This motivates the cold ion limit that underlies the HM equation, which eliminates the inconvenient terms, but is however not generally appropriate for application to fusion plasmas.  In the present work, it is noted that the presence of these terms introduces rapid dynamics, and a multiple-scales analysis is proposed in which the fluid moment hierarchy closes at the pressure moment, without using an ad hoc closure scheme, leading to a relatively simple system involving only two fields.

What is immediately apparent is that the presence of the additional field (the pressure perturbation) breaks the nonlinear conservation of enstrophy that is famously satisfied by the HM equation, and there causes an ``inverse cascade'' of energy to large scales.  The new system, we argue, should exhibit distinct nonlinear behavior, including a shallower energy spectrum when the effect of the nonlinear FLR terms is sufficiently strong.  Direct numerical simulation of the fluid model gives some confidence in these predictions.  The results of this work may help to interpret observations of turbulence in parameter regimes where the dynamics tend toward the quasi-two-dimensional limit.  We discuss possible examples, including cases explored in previous gyrokinetic turbulence simulations in tokamak and stellarator geometries.

\section{Equations and definitions}\label{eqns-sec}

We assume uniform magnetic geometry, where the magnetic guide field is constant and points in the $\hat{z}$-direction.  One species is assumed to be kinetic, with the other species satisfying a simple Boltzmann response model.  We begin with a nondimensional form of the gyrokinetic system \citep{plunk-jfm}, normalized relative to the kinetic species: $v_{\perp}/\vth \rightarrow v$ (with $\vth = \sqrt{T/m}$, and $T$ and $m$ are the temperature and mass of the kinetic species) is the normalized perpendicular velocity and the normalized wavenumber is $k_{\perp}\rho \rightarrow k$ where thermal Larmor radius of the kinetic species is $\rho = \vth/\Omega_c$ and $\Omega_c = qB/m$.  The two-dimensional gyrokinetic equation is written as follows in terms of the perturbed gyrocenter distribution function $g({\bf R}, v, t)$, where ${\bf R} = \hat{\bf x} X + \hat{\bf y} Y$ is the gyrocenter position:

\begin{equation}
\frac{\partial g}{\partial t} +  \poiss{\gyroavg{\varphi}}{g} = \CollisionOp{h}.
\label{gyro-g}
\end{equation}

\noindent where $\CollisionOp{h}$ is the collision operator (not treated here explicitly); the Poisson bracket is $\poiss{A}{B} = \hat{\bf z}\times\bnabla A \cdot \bnabla B = \partial_x A \partial_y B - \partial_y A \partial_x B$ and the gyro-average is defined $\gyroavg{A({\bf r})} = \frac{1}{2\pi}\int_0^{2\pi} d\vartheta A({\bf R} + \rhovec(\vartheta))$, where the Larmor radius vector is $\rhovec(\vartheta) = {\bf {\hat z}}\times{\bf v} = v_{\perp}({\bf {\hat y}}\cos{\vartheta} - {\bf {\hat x}}\sin{\vartheta})$ and $\vartheta$ is the gyro-angle.  (Note that the quantity inside of the collision operator is $h = g + \gyroavg{\varphi}F_0$.  Note also that the spatial coordinate is ${\bf R}$ in the gyrokinetic equation and, formally, the spatial derivatives are to be interpreted in this variable, but for simplicity we avoid making the distinction explicit.)  We mostly ignore the collision operator but note that some mechanism of coarse-graining will be necessary to get sensible solutions out of the equation.  Quasi-neutrality yields the electrostatic potential $\varphi({\bf r}, t)$, where ${\bf r} = \hat{\bf x} x + \hat{\bf y} y$ is the position-space coordinate:

\begin{equation}
2\pi\int_0^{\infty} v dv \angleavg{g} = (1 + \tau)\varphi - \Gamma_0\varphi,
\label{qn-g}
\end{equation}

\noindent where the $g$ is implicitly assumed to be integrated over parallel velocity so that $2\pi\int_0^{\infty} v dv$ completes the integration over three-dimensional velocity space.  The angle average is defined $\angleavg{A({\bf R})} = \frac{1}{2\pi}\int_0^{2\pi} d\vartheta A({\bf r} - \rhovec(\vartheta))$, and the term $\tau\varphi$ is the adiabatic density response, and $\tau = T_i/(Z T_e)$ for the case of ion scales and $\tau = Z T_e/ T_i$ for the case of electron scales.  For the ion case, this Boltzmann response might be considered reasonable if zonal flows are strongly suppressed.  The operator $\Gamma_0\phi = 2\pi\int_0^{\infty} v dv \;F_0(v)\angleavg{\gyroavg{\phi}}$ is more naturally expressed in Fourier space, assuming a Maxwellian background $F_0 = \exp[-v^2/2]/(2\pi)$, \ie $\Gamma_0\varphi = \sum_{\bf k} \exp(i {\bf k}\cdot{\bf r}) \hat{\Gamma}_0 \hat{\varphi}$, with

\begin{equation}
\hat{\Gamma}_0(k) = \int_0^{\infty} v dv \;\e^{-v^2/2}J_0^2(kv) = I_0(k^2)e^{-k^2},\label{gamma0-def}
\end{equation}

\noindent where $I_0$ is the zeroth-order modified Bessel function.
\section{Fluid limit}\label{eqns-sec}

We expand in the limit

\begin{equation}
\delta = k^2 \ll 1,
\end{equation}
\ie we assume that the scales of interest are larger than the Larmor radius of the species of interest.  For electron scales, the limit is considered subsidiary to the adiabatic ion limit, so scales of the turbulence must remain much smaller than the ion Larmor scale, \ie, $\rho_e/\rho_i \ll k \ll 1$.  Note that there may also be a minimum applicable $k$ imposed by dynamics parallel to the magnetic field, but treating this explicitly is outside the scope of this work.  We will only need the first two orders of the expansion in $\delta$. The gyrokinetic equation, henceforth omitting explicit collisional effects, is written as

\begin{equation}
\frac{\partial g}{\partial t} + \poiss{\left(1+ \frac{v^2}{4} \nabla^2\right)\varphi}{g} \approx 0,\label{gk-lw-eqn}
\end{equation}

\noindent and Eqn.~\ref{qn-g} becomes

\begin{equation}
\tau\varphi - \nabla^2\varphi \approx 2\pi \int_0^{\infty} vdv \left(1+ \frac{v^2}{4} \nabla^2\right) g.\label{qn-lw-eqn}
\end{equation}
We will denote $v$-moments of $g$ as 

\begin{equation}
G_n = 2\pi \int vdv \left(\frac{v}{2}\right)^n g.
\end{equation}

\subsection{Naive expansion}\label{naive-sec}
To give a taste for the issues that arise in the expansion, let us take an initial informal look at the moments of the gyrokinetic equation.  We first examine the density moment.  We include only the ostensibly dominant nonlinear terms.  We stress that this equation is given only for illustrative purposes, and is not to be taken as a basis for the later derivations of the paper:
\begin{equation}
\frac{\partial}{\partial t}\left(\tau \varphi -\nabla^2\varphi - \nabla^2 G_2\right) + \poiss{\varphi}{-\nabla^2 \varphi -\nabla^2 G_2 } + \poiss{G_2}{ -\nabla^2 \varphi} = 0.\nonumber
\end{equation}
Note that the term $-\partial_t \nabla^2 \varphi$, which appears in the HM equation, should be neglected here because it is formally smaller than $\partial_t \varphi$ by one power of the ordering parameter $\delta$.  Likewise, the term $-\partial_t \nabla^2 G_2$ must be considered negligible if the ordering $G_2 \sim \varphi$ and $\partial_t G_2 \sim \partial_t \varphi$ holds.  The situation is, however, a bit more subtle.  The above equation couples to the $v^2$ moment of $g$, $G_2$, and the equation for this and other such moments can be written, neglecting higher-order FLR terms, as

\begin{equation}
\frac{\partial G_n}{\partial t} + \poiss{\varphi}{G_n} = 0.\label{Gn-nonresonant-eqn}
\end{equation}

What we now notice, examining these two equations, is that the density equation is driven by nonlinear terms that appear to be much smaller than those controlling the higher moments of the distribution function -- that is, the equations for $G_n$ have dominant contributions from the $E\times B$ nonlinearity, while the potential evolves under the influence of terms like the ``polarization drift'' nonlinearity, which is smaller by a factor of $\delta = k^2$. One possible resolution of this apparent imbalance is to consider $G_n$ itself to be large, as for instance in the non-resonant limit of the ITG/ETG mode, \ie $G_n \sim \delta^{-1} \varphi$.  In this case, the polarization drift nonlinearity can be neglected, and we see the justification for retaining the additional time derivative term above, since $\partial_t\varphi \sim \partial_t \nabla^2 G_2$.  This term can be evaluated from the Laplacian of Eqn.~\ref{Gn-nonresonant-eqn}, yielding

\begin{equation}
\tau \frac{\partial \varphi}{\partial t} + \nabla^2\poiss{\varphi}{G_2} + \poiss{\varphi}{-\nabla^2 G_2 } + \poiss{G_2}{ -\nabla^2 \varphi} = 0.\label{phi-nonresonant-eqn}
\end{equation}

Eqns.~\ref{Gn-nonresonant-eqn}-\ref{phi-nonresonant-eqn} demonstrate a consistent fluid limit, but cannot describe ITG or ETG turbulence in the resonant limit, where $\varphi \sim G_2$.  This corresponds the more physically reasonable scenario of a modest turbulence drive -- \ie not very far from the linear critical gradient, or considering the weakly unstable, large-scale modes that dominate the turbulence spectrum.  To treat this limit properly, we must account for the fact that $\varphi$ evolves much more slowly than $G_n$.  Physically, it can be argued that, in a turbulent state, Eqn.~\ref{Gn-nonresonant-eqn} will then describe rapid mixing of $G_n$ by $E\times B$ vortices, so that any initial variation along streamlines of constant $\varphi$ will decay on a fast timescale (with the help of some explicit dissipation), leaving $G_n$ to be constant along those streamlines \citep{cowley-private}.  To account for such processes more carefully, we abandon conventional perturbation theory in favor of the method of multiple scales (see, \ie \cite{bender}).  We will henceforth disregard the equations presented here, in section \ref{naive-sec}, and proceed to derive equations that contain only terms justified by a set of explicitly stated ordering assumptions.

\subsection{Multiscale expansion}
We introduce the fast and slow time variables $t_\mathrm{f}$, and $t_\mathrm{s}$, such that $\partial_{t_\mathrm{f}} \sim \poiss{\varphi}{ .}$ and $\partial_{t_\mathrm{s}} \sim \poiss{\nabla^2\varphi}{ .} \sim \delta \partial_{t_\mathrm{f}}$ and expand the fields as

\begin{eqnarray}
\varphi = \varphi^{(0)}(t_\mathrm{s}, t_\mathrm{f}, x, y) + \varphi^{(1)}(t_\mathrm{s}, t_\mathrm{f}, x, y) + \dots,\\
G_n = G_n^{(0)}(t_\mathrm{s}, t_\mathrm{f}, x, y) + G_n^{(1)}(t_\mathrm{s}, t_\mathrm{f}, x, y) + \dots,
\end{eqnarray}
where $\varphi^{(m+1)}/\varphi^{(m)} \sim {\cal O}(\delta)$, {\em etc}.  We reiterate that the assumptions we have made are $\delta \ll 1$, the above multi-scale expansion, and the quasi-two-dimensional approximation, whereby variation in the direction along the magnetic field is neglected, and the non-kinetic species is assumed to follow a Boltzmann distribution, implying Eqn.~\ref{qn-g}; no further approximations will be made in this section.  We proceed to examine the moments of gyrokinetic equation, order by order.  The density moment at dominant order in $\delta$ is

\begin{equation}
\frac{\partial \varphi^{(0)}}{\partial t_\mathrm{f}} = 0,
\end{equation}
from which we formally establish that $\varphi^{(0)}$ depends only on the slow time variable.  At next order in $\delta$ we obtain

\begin{equation}
\tau \frac{\partial \varphi^{(0)}}{\partial t_\mathrm{s}} + \tau\frac{\partial \varphi^{(1)}}{\partial t_\mathrm{f}} - \frac{\partial}{\partial t_\mathrm{f}}\nabla^2 G_2^{(0)} + \poiss{\varphi^{(0)}}{-\nabla^2 \varphi^{(0)} -\nabla^2 G_2^{(0)} } + \poiss{G_2^{(0)}}{ -\nabla^2 \varphi^{(0)}} = 0.\label{phi-0-eqn}
\end{equation}
We introduce a time-average operator to extract the smooth-time behavior from this equation

\begin{equation}
\tAvg{A} = \frac{1}{\Delta t} \int_{t_\mathrm{f}-\Delta t/2}^{t_\mathrm{f}+\Delta t/2} dt_\mathrm{f}^\prime A(t_\mathrm{f}^\prime).\label{tAvg-def}
\end{equation}
This time average extends over a period of time much longer than the short timescale ($\Delta t^{-1} \ll \poiss{\varphi}{.}$).  Applying this average to Eqn.~\ref{phi-0-eqn}, we obtain

\begin{equation}
\tau \frac{\partial \varphi^{(0)}}{\partial t_\mathrm{s}} + \poiss{\varphi^{(0)}}{-\nabla^2 \varphi^{(0)} -\nabla^2 \tAvg{G}_2^{(0)} } + \poiss{\tAvg{G}_2^{(0)}}{ -\nabla^2 \varphi^{(0)}} = 0.\label{phi-0-avg-eqn}
\end{equation}
The dominant-order equation for $G_n$ is

\begin{equation}
\frac{\partial G_n^{(0)}}{\partial t_\mathrm{f}} + \poiss{\varphi}{G_n^{(0)}} = 0,\label{G0-fast-eqn}
\end{equation}
from which, upon time averaging, we conclude that $\tAvg{G}_n^{(0)}$ is constant along closed streamlines of constant $\varphi$.  Informally, we will say that $\tAvg{G}_n^{(0)}$ is a function of $\varphi$, although it can be multi-valued.  For $n = 2$ we adopt the notation

\begin{equation}
\tAvg{G}_2^{(0)} = \chi(\varphi, t_\mathrm{s}).
\end{equation}
Note that, formally, we must exclude special points and lines where $\bnabla \varphi = 0$ (o-points, and the ``separatrices'' that include x-points) but these should occupy negligible volume in the $x$-$y$ plane.  At the next order, we will obtain the smooth evolution of $G_n$,

\begin{equation}
\frac{\partial G_n^{(0)}}{\partial t_\mathrm{s}} + \frac{\partial G_n^{(1)}}{\partial t_\mathrm{f}} + \poiss{\varphi^{(0)}}{G_n^{(1)}} + \poiss{\varphi^{(1)}}{G_n^{(0)}} + \poiss{\nabla^2 \varphi^{(0)}}{G_{n+ 2}^{(0)}} = 0.\label{G0-eqn}
\end{equation}
To this equation we apply two averages, the time average, and also an average along streamlines of constant $\varphi$.  To define this average, we introduce a coordinate $s$ which parameterizes these streamlines and satisfies $\hat{\bf z}\times\bnabla \varphi \cdot \bnabla s = 1$ for convenience.  Then we define

\begin{equation}
\sAvg{A} = \frac{\oint ds A(s)}{\oint ds}.
\end{equation}
The integral over $s$ is either closed in the sense that the streamlines are closed, or effectively closed by periodic boundary conditions.  The second term of Eqn.~\ref{G0-eqn} is annihilated by the time average.  Noting that $\poiss{F(\varphi)}{ A} = \partial_s(A F')$, the third term is zero under the $s$-average, as is the last term after time average, since $\tAvg{G}_{n + 2}^{(0)}$ is a function of $\varphi^{(0)}$ by Eqn.~\ref{G0-fast-eqn}.  This is a crucial cancellation since the fluid moment hierarchy is consequently shown to be closed.

It is convenient to now introduce notation for the part of a field that varies on the fast timescale, \ie the ``fluctuating part'', complementing the mean component defined by Eqn.~\ref{tAvg-def}:

\begin{equation}
\nAvg{A} = A - \tAvg{A}.\label{nAvg-def}
\end{equation}
What results from the double average of Eqn.~\ref{G0-eqn} can then be expressed

\begin{equation}
\sAvg{\frac{\partial \tAvg{G}_n^{(0)}}{\partial t_\mathrm{s}}} + \sAvg{\tAvg{\poiss{\nAvg{\varphi}^{(1)}}{\nAvg{G}_n^{(0)}}}} = 0.\label{G0-avg-eqn}
\end{equation}
To evaluate the nonlinear term of Eqn.~\ref{G0-avg-eqn}, we must obtain dynamical equations for the fluctuating fields $\nAvg{\varphi}^{(1)}$ and $\nAvg{G}_n^{(0)}$.  These come from Eqns.~\ref{phi-0-eqn} and \ref{G0-fast-eqn}, respectively.  The fluctuating part of Eqn.~\ref{G0-fast-eqn} is

\begin{equation}
\frac{\partial \nAvg{G}_n^{(0)}}{\partial t_\mathrm{f}} + \poiss{\varphi^{(0)}}{\nAvg{G}_n^{(0)}} = 0.\label{G0-fast-eqn-final}
\end{equation}
Subtracting Eqn.~\ref{phi-0-avg-eqn} from Eqn.~\ref{phi-0-eqn}, and using the Laplacian of Eqn.~\ref{G0-fast-eqn-final} to evaluate $\partial_{t_\mathrm{f}}\nabla^2 \nAvg{G}_n^{(0)}$, we find

\begin{equation}
\tau\frac{\partial \nAvg{\varphi}^{(1)}}{\partial t_\mathrm{f}}  + \nabla^2\poiss{\varphi^{(0)}}{\nAvg{G}_2^{(0)}} + \poiss{\varphi^{(0)}}{-\nabla^2 \nAvg{G}_2^{(0)} } + \poiss{\nAvg{G}_2^{(0)}}{ -\nabla^2 \varphi^{(0)}} = 0.\label{phi-1-navg-eqn}
\end{equation}
Finally, noting that $\sAvg{\partial_{t_\mathrm{s}}\varphi^{(0)}} = 0$ (from Eqn.~\ref{phi-0-avg-eqn}), we obtain, from Eqn.~\ref{G0-avg-eqn}, an expression determining the explicit time dependence of $\chi$:

\begin{equation}
\left(\frac{\partial \chi}{\partial t_\mathrm{s}}\right)_{\varphi} + \sAvg{\tAvg{\poiss{\nAvg{\varphi}^{(1)}}{\nAvg{G}_2^{(0)}}}} = 0,\label{G0-avg-eqn-2}
\end{equation}
where the partial time derivative is taken at constant $\varphi^{(0)}$.  To summarize, Eqns.~\ref{G0-fast-eqn-final} and \ref{phi-1-navg-eqn} are the fast-time equations that determine $\nAvg{\varphi}^{(1)}$ and $\nAvg{G}_n^{(0)}$, which can be substituted into the slow-time equation \ref{G0-avg-eqn-2} for $\chi$, and coupled with the following equation (a repetition of Eqn.~\ref{phi-0-avg-eqn} written in terms of $\chi$) to close the system:

\begin{equation}
\tau \frac{\partial \varphi^{(0)}}{\partial t_\mathrm{s}} + \poiss{\varphi^{(0)}}{-\nabla^2 \varphi^{(0)} -\nabla^2 \chi } + \poiss{\chi}{ -\nabla^2 \varphi^{(0)}} = 0.\label{phi-0-avg-eqn-2}
\end{equation}
Noting that $\tAvg{\varphi} \approx \varphi^{(0)}$ and $\nAvg{\varphi} \approx \nAvg{\varphi}^{(1)}$, we can, without introducing ambiguity, simply drop the superscripts in what follows.  

The final system, Eqns.~\ref{G0-fast-eqn-final}-\ref{phi-0-avg-eqn-2}, has some noteworthy features.  First, Eqn.~\ref{G0-avg-eqn-2} has the appearance of a heat transport equation, where the flux is carried by the rapidly varying pressure perturbation $\nAvg{G}_2$ and the small amplitude fluctuating potential $\nAvg{\varphi}$.  Note also how Eqns.~\ref{G0-fast-eqn-final} and \ref{phi-1-navg-eqn} bear a strong resemblance to the fluid system given by Eqns.~\ref{Gn-nonresonant-eqn}-\ref{phi-nonresonant-eqn}, where a similar ordering is satisfied, namely $\nAvg{\varphi} \ll \nAvg{G}_2$.  

\section{Decaying turbulence}

We will avoid the complications introduced by the instabilities that physically drive turbulence, and instead now consider decaying turbulence.  (We note that a linear instability could be added to this fluid system using the non-resonant limit of the toroidal branch of the ITG or ETG mode, but this would require some care to maintain consistency with the ordering assumptions, as discussed in section \ref{naive-sec}.)   Let us consider periodic boundary conditions, and include explicit dissipation using a fourth-order hyperviscosity term.  Without drive terms, Eqn.~\ref{G0-fast-eqn-final} implies the rapid decay of $\nAvg{G}_n$ to zero, implying  $(\partial_{t_\mathrm{s}}\chi)_{\varphi} = 0$ (\ie it only depends on the time via its dependence on $\varphi$).  The variation of $\chi$ in $\varphi$ (or more formally, its variation between distinct lines of constant $\varphi$) can be considered as an initial condition of our calculation.  We need only then solve a single equation, which, neglecting superscripts for order and the subscripts of the time variable $t_\mathrm{s}$, becomes

\begin{equation}
\tau \frac{\partial \varphi}{\partial t} + \poiss{\varphi}{-\nabla^2 \varphi -\nabla^2 \chi } + \poiss{\chi}{ -\nabla^2 \varphi} = \nu_4 \nabla^4 \varphi.\label{phi-chi-eqn}
\end{equation}

The electrostatic energy

\begin{equation}
E = \frac{\tau}{2}\int dx dy \varphi^2
\end{equation}
is conserved by the nonlinearity for arbitrary $\chi$, which can be verified by multiplying the equation by $\varphi$ and integrating over the $x$-$y$ domain.  Note that the resulting integral of the final nonlinear term of Eqn.~\ref{phi-chi-eqn} can be rewritten as $- \int dx dy\; {\bf v}_E\cdot\bnabla(\varphi \chi^\prime \nabla^2\varphi)$, with ${\bf v}_E = \hat{\bf z}\times\bnabla\varphi$, which is zero using $\bnabla\cdot{\bf v}_E = 0$ and periodicity.

Another quantity of interest is the enstrophy, which we will define here as

\begin{equation}
Z = \frac{\tau}{2}\int dx dy |\bnabla\varphi|^2.
\end{equation}
The enstrophy balance equation is found by multiplying Eqn.~\ref{phi-chi-eqn} by $-\nabla^2\varphi$ and integrating over $x$ and $y$.  Note that the presence of $\chi$ in the equation breaks enstrophy conservation if $\chi$ is a nonlinear function of $\varphi$.  The nonlinear invariance of $Z$ is associated with the inverse cascade of energy in Hasegawa-Mima turbulence.  We thus expect to recover the spectra corresponding to the potential limit of the Hasegawa-Mima equation (\ie where $\nabla^2 \varphi \ll \varphi$; see \citet{plunk-jfm}), if $\chi$ is small, and qualitatively different cascade when $\chi$ is sufficiently large.  

The Hasegawa-Mima spectra can be derived in the rough ``phenomenological'' style, in terms of the fluctuation amplitude at scale $\ell$, denoted $\varphi_\ell$, by assuming constancy of nonlinear flux of its nonlinear invariants (see \eg \citet{frisch, plunk-jfm}).  For the forward cascade, \ie at scales smaller than the scale of energy injection, the enstrophy flux, denoted $\varepsilon_Z$, is assumed constant (independent of scale $\ell$), which is expressed as follows:

\begin{equation}
\varepsilon_Z = \tau_{\mathrm{NL}}^{-1} \ell^{-2} \varphi_\ell^2 \sim \varphi_\ell^3\ell^{-6},
\end{equation}
with $\tau_{\mathrm{NL}}(\ell)$ denoting the nonlinear turnover time.  This leads to the scaling $\varphi_\ell \sim \ell^2\varepsilon_Z^{1/3}$, implying a one-dimensional energy spectrum of $E(k) \sim k^{-5}$.  The constancy of the scale-by-scale flux of energy, expected for the inverse cascade at scales larger than the injection scale, is expressed as
\begin{equation}
\varepsilon_E = \tau_{\mathrm{NL}}^{-1} \varphi_\ell^2 \sim \varphi_\ell^3\ell^{-4},
\end{equation}
implying $\varphi_\ell \sim \ell^{4/3}\varepsilon_E^{1/3}$ and a spectrum $E(k) \sim k^{-11/3}$.  

Because the additional nonlinear terms of Eqn.~\ref{phi-chi-eqn}, henceforth called the ``$\chi$ nonlinearity'', formally break enstrophy conservation, we expect that if they are sufficiently strong, the inverse cascade should be eliminated and the forward cascade of $Z$ replaced with a direct cascade of $E$.  If this flux is carried by the HM nonlinearity, one might expect to observe the spectrum $E(k) \sim k^{-11/3}$, as suggested by \citet{plunk-prl-2019}.  On the other hand, balancing the $\chi$ nonlinearity with the HM nonlinearity, scale-by-scale, implies the linear relation $\chi_\ell \sim \varphi_\ell$, \ie $\chi \propto \varphi$, which would imply that enstrophy is actually a nonlinear invariant, preventing the forward cascade of $E$.  For this reason, we may expect to observe an energy spectrum distinct from $k^{-11/3}$, whose steepness depends on the relationship between $\chi_\ell$ and $\varphi_\ell$, which itself depends on details of the turbulence.

Providing a definitive prediction of this relationship is beyond the scope of the present work, but a power law seems to be a reasonable possibility to explore, \ie $\chi_\ell \sim \varphi_\ell^\alpha$.  Note that any super-linear scaling $\alpha > 1$ should lead to a spectrum shallower than $k^{-11/3}$, while a sub-linear scaling $\alpha < 1$ would imply $\chi$ is not analytic in $x$ and $y$.  The fluctuating fields $\nAvg{G}_2$ and $\nAvg{\varphi}$ could be especially active in regions of low $E\times B$ shear (see Eqn.~\ref{G0-fast-eqn-final}), causing local extrema in the function $\chi$, via Eqn.~\ref{G0-avg-eqn-2}, so that a quadratic relationship prevails in such regions, $\chi_\ell \sim \varphi_\ell^2$.  Whether or not this seems plausible, assuming a simple nonlinear relationship will allow us to make the discussion now more concrete; qualitatively similar conclusions should apply for all $\alpha > 1$.  Let us consider the following form for $\chi$:

\begin{equation}
\chi(\varphi) = \frac{\lambda}{2} \varphi^2.\label{chi-eqn}
\end{equation}
The nonlinear energy flux by the $\chi$ terms is then expressed as $\varepsilon_E \sim \lambda \varphi_\ell^4 \ell^{-4}$, implying $\varphi_\ell \sim \ell (\varepsilon_E/\lambda)^{1/4}$, and the corresponding energy spectrum

\begin{equation}
E(k) \sim k^{-3}.
\end{equation}
This spectrum should prevail in cases where the $\chi$-nonlinearity dominates (\eg large $\lambda$).  At sufficiently low $\lambda$, one expects a return to the HM behavior, implying $E(k) \sim k^{-5}$ for the forward cascade.  

Some sort of hybrid behavior may also be possible, although the broad scale range needed for clear observation of this may be not be present for realistic conditions encountered in fusion plasmas.  One might argue that, because the amplitude of fluctuations $\varphi_\ell$ is generally expected to decrease as scales do, the cubic nonlinearity should be dominant at large scales, and subdominant at small scales.  Thus, for sufficiently large $\lambda$, the energy cascade scaling $\varphi_\ell \sim \ell (\varepsilon_E/\lambda)^{1/4}$ should hold from the injection scale, down to a transition scale, which can be found by balancing the HM nonlinearity with the $\chi$-nonlinearity, \ie $\varphi_\ell^2\ell^{-4} \sim \lambda \varphi_\ell^3 \ell^{-4}$.  Defining the outer scale $\ell_\mathrm{o}$ as the scale of energy injection (or initial energy containing scale), and $\varphi_{\mathrm{o}} = \varphi_{\ell_\mathrm{o}}$, we can write the $\varphi_\ell$ scaling as $\varphi_\ell \sim (\ell/\ell_\mathrm{o}) \varphi_{\mathrm{o}}$, so that the above balance occurs at the ``transition'' scale $\ell_t \sim \ell_\mathrm{o} / (\lambda \varphi_{\mathrm{o}})$.  Thus, if $\lambda \varphi_{\mathrm{o}} \gtrsim 1$ one might expect $E \sim k^{-3}$ scaling for $\ell_{\mathrm{o}}^{-1} < k < \ell_t^{-1}$ followed by $E \sim k^{-5}$ for $k > \ell_t^{-1}$.

\subsection{Direct numerical simulations}

To explore the behavior of the model, Eqn.~\ref{phi-chi-eqn}, and test the theoretical predictions, we perform direct numerical simulations, assuming the simple quadratic form of $\chi(\varphi)$ in Eqn.~\ref{chi-eqn}.  This introduces a nonlinearity that is cubic in $\varphi$, which can be treated pseudo-spectrally using a padding factor of $2$ for dealiasing; higher order nonlinearities require additional padding \citep{HOSSAIN1992}.  The boundary conditions for the simulations are periodic in $x$ and $y$, and $\tau = 1$ for all simulations.

\begin{figure}
\includegraphics[width=0.45\columnwidth]{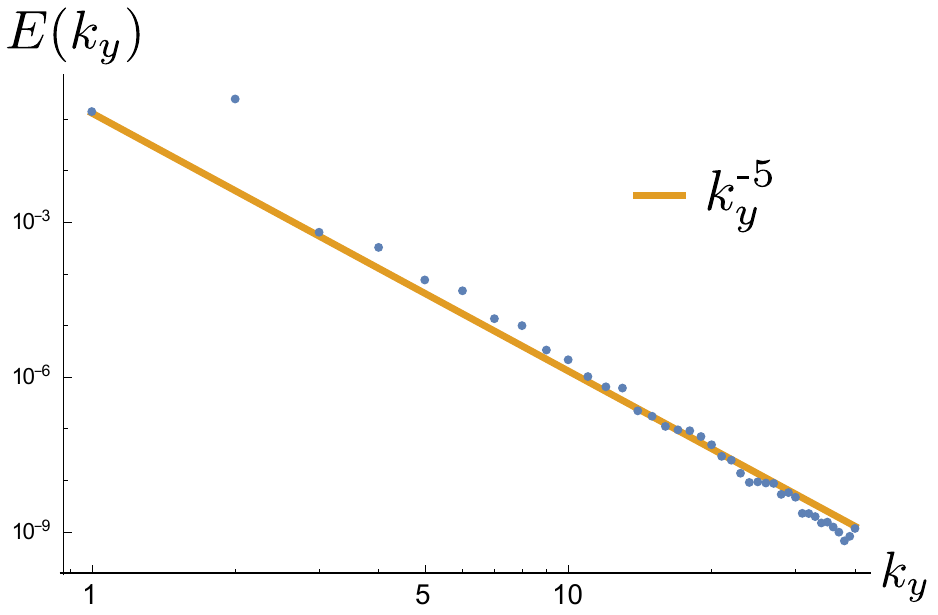}
\includegraphics[width=0.45\columnwidth]{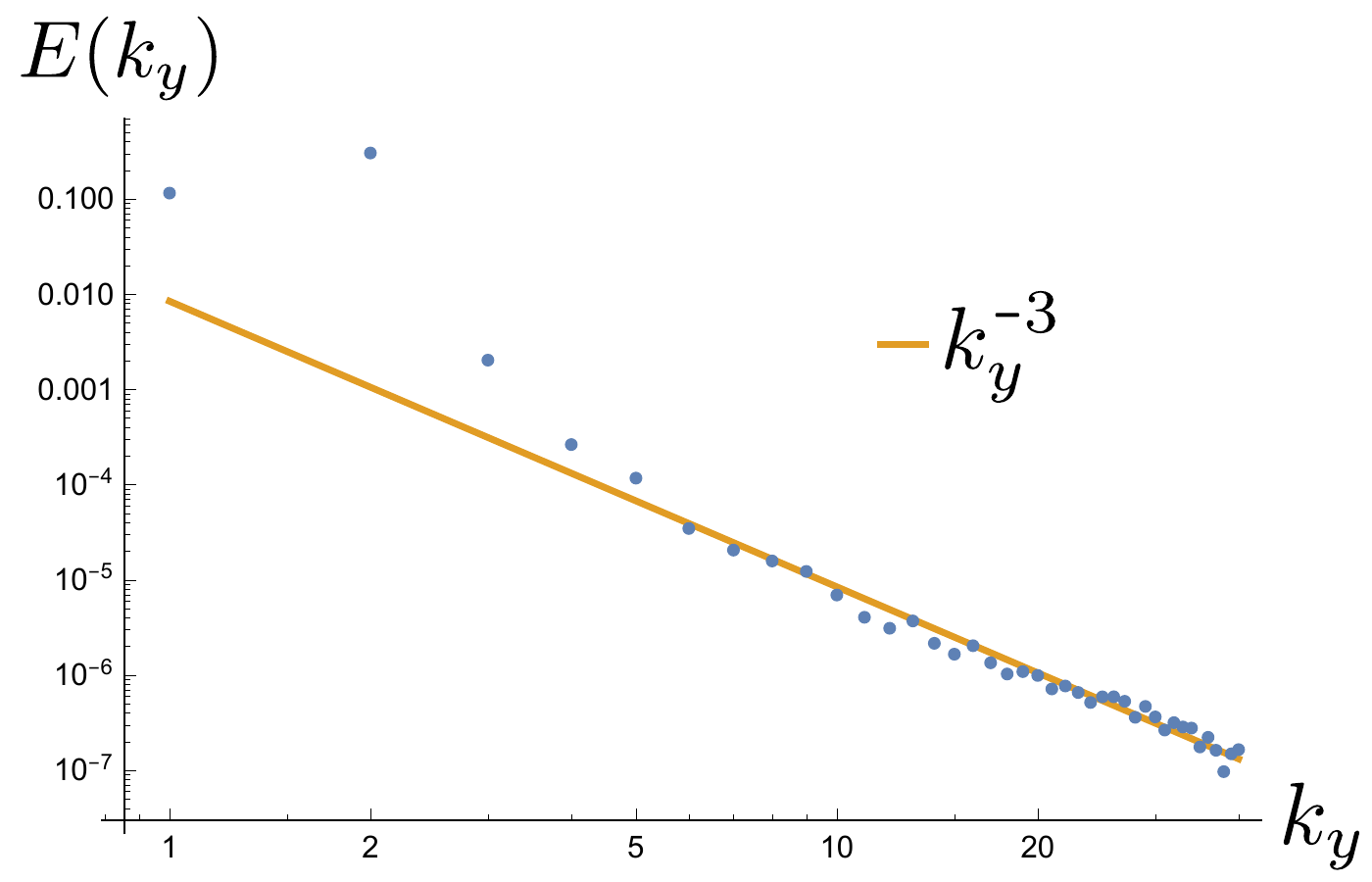}
\caption{Comparison of spectra exhibited by HM-type system ($\lambda = 0$) and our two-dimensional turbulence model ($\lambda \rightarrow \infty$).}
\label{E-spectra-fig}
\end{figure}

Fig.~\ref{E-spectra-fig} compares the simulation results with the theoretical scaling laws.  All simulations are initialized with randomly phased fluctuation amplitude of $\varphi \sim 1$ around $k = 1$, falling off exponentially at higher $k$.  Note that although the model assumes $k \ll 1$ there is no conflict in using $k > 1$ for the simulations, as scaling symmetries of the model allow the results to be reinterpreted for $k \ll 1$.  The spectrum found for the $\lambda = 0$ case is roughly consistent with the theoretical power law $k^{-5}$ expected for the potential limit of the HM equation.  We note that similar results (not shown here) are encountered for $\lambda \lesssim 0.1$.  At larger $\lambda$, the breaking of enstrophy conservation is indeed observed in the time trace of $Z$, as the energy fills in the spectrum at large $k$.  For the case labeled $\lambda \rightarrow \infty$ in Fig.~\ref{E-spectra-fig}, the spectrum seems consistent with the theoretical $k^{-3}$ prediction at scales smaller than the injection scale.  Note that this limit is obtained by actually setting $\lambda = 1$ and simply removing the HM nonlinearity (\ie the first term of Eqn.~\ref{phi-chi-eqn}) from the equation, as can be formally justified by rescaling Eqn.~\ref{phi-chi-eqn} in the limit $\lambda \rightarrow \infty$.  Similar behavior is observed for $\lambda \gtrsim 1$.  Intermediate values of $\lambda$ show intermediate behavior.  

\begin{figure}
\begin{center}
\includegraphics[width=0.45\columnwidth]{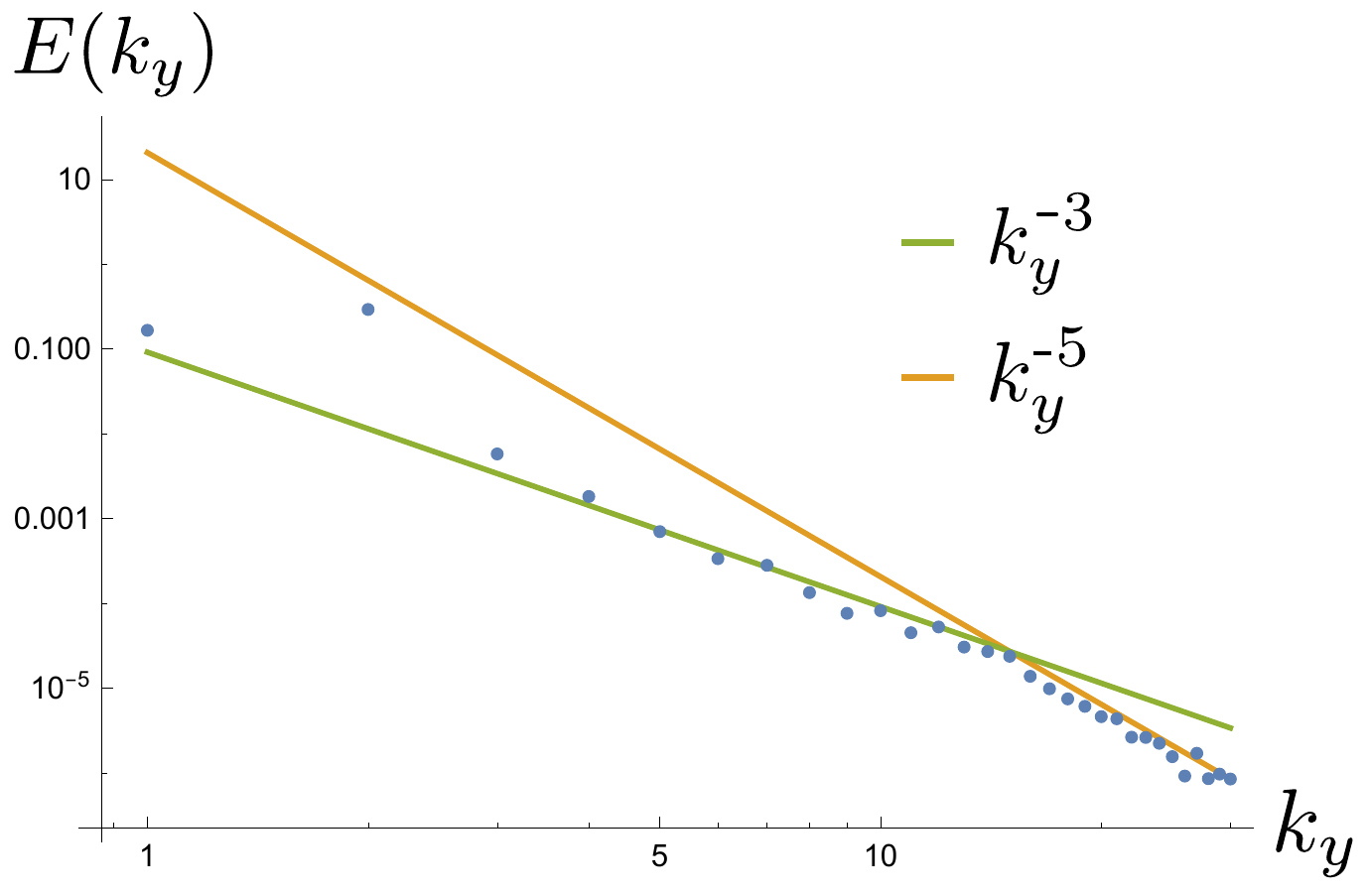}
\end{center}
\caption{Energy spectrum for a case of intermediate strength of $\chi$ nonlinearity ($\lambda = 0.5$).}
\label{lambda-mid-fig}
\end{figure}

One example is shown in Fig.~\ref{lambda-mid-fig}, which seems to show evidence of a transition scale between the two theoretical power laws, giving some support to the predictions of a hybrid scenario described theoretically in the previous section.  A more extensive set of simulations would be needed to test the predictions in detail, for instance the dependence of the transition scale $\ell_t$ on system parameters.  We would like to generally stress that the results of the numerical simulations presented here come at a very modest computational expense, and larger scale computational effort, especially using a gyrokinetic code, could offer a more extensive test of the conclusions of this work.

\section{Discussion}

A novel fluid system has been derived to describe the behavior of certain classes of quasi-two-dimensional electrostatic magnetized plasma turbulence.  A possible application is to describe the energy cascade in cases of streamer-dominated ETG turbulence (note the spectrum, noted to be close to $k^{-11/3}$, in Fig.~5 of \citet{plunk-prl-2019}), where the nonlinear stability of elongated turbulent eddies is believed to stem from the two-dimensional character of the dominant instabilities, \eg the absence of sufficient variation of the mode structure in the direction along the magnetic field \citep{jenko-dorland-prl-2002}.  This turbulent state is, however, sensitive to magnetic geometry, and seems to vanish when, for instance, the global magnetic shear is varied in such a way as to induce stronger parallel electron flow to the ETG mode.  The ensuing dynamics then depends on kinetic physics involving the parallel streaming term, absent from two-dimensional models.  A second possible application of the present model might be to describe ITG turbulence in cases where the zonal flows are suppressed.  One candidate is a case observed with simulations of the HSX stellarator having surprisingly steep fluctuation spectra \citep{plunk-prl-2017}, found to be close to $k^{-10/3}$.

Although the presented model has limited application, it fills a significant gap in present theories describing gyrokinetic turbulence cascades, as it accounts for the essential nonlinear terms that arise when the cold ion approximation is invalid.  These terms, it is found, alter the conservative properties of the nonlinearity, with significant consequences on the cascade, so that, even in the two-dimensional limit, the inverse cascade of energy can be shut down.  The numerical simulations confirm that the size of the pressure perturbation ($\chi$) can control the cascade type, and HM-like behavior can be recovered if it is sufficiently small.  This may underlie the slow secular growth of large-scale zonal flows \citep{Guttenfelder-Candy} and other coherent structures \citep{nakata-vortex-streets} in simulations of ETG turbulence, and the related appearance of a Dimits shift phenomenon in near-marginal cases \citep{Colyer_2017}.

{\bf Acknowledgements.}  This work has been carried out within the framework of the EUROfusion Consortium and has received funding from the Euratom research and training programme 2014-2018 and 2019-2020 under grant agreement No 633053. The views and opinions expressed herein do not necessarily reflect those of the European Commission.

\bibliographystyle{unsrtnat}
\bibliography{2D-electron-turbulence-model}
\end{document}